\documentclass[showpacs,preprintnumbers]{revtex4}
\usepackage{amsmath}
\usepackage{graphicx}

\begin{document}

\textbf{Simple Proof that the Usual Transformations of the Electric and }

\textbf{Magnetic Fields are not the Lorentz Transformations and EDM Searches
}\bigskip

Tomislav Ivezi\'{c}

\textit{Ru%
\mbox
{\it{d}\hspace{-.15em}\rule[1.25ex]{.2em}{.04ex}\hspace{-.05em}}er Bo\v
{s}kovi\'{c} Institute, P.O.B. 180, 10002 Zagreb, Croatia}

\textit{ivezic@irb.hr\bigskip }

The usual transformations of the three-dimensional (3D) fields $\mathbf{E}$
and $\mathbf{B}$ that are found in [1] ([1] A. Einstein, Ann. Physik \textbf{%
17}, 891 (1905)) are always considered to be the relativistically correct
Lorentz transformations (LT) of $\mathbf{E}$ and $\mathbf{B}$. However, as
proved in, e.g., [2] ([2] T. Ivezi\'{c}, Found. Phys. Lett. \textbf{18, }301
(2005)), these transformations drastically differ from the LT of the
relativistically correct 4D electric and magnetic fields. In this paper a
simple proof of that difference will be presented and the consequences for
EDM experiments and for some quantum phase shifts experiments are briefly
examined. In all such experiments the usual 3D quantities, e.g., $\mathbf{E}$%
, $\mathbf{B}$, \textbf{...} are measured and their relativistically
incorrect transformations are used, but not the relativistically correct 4D
geometric quantities, e.g., $E^{a}$, $B^{a}$, ... and their LT.\bigskip
\medskip

\medskip
\bigskip

\textit{Introduction.} - It is generally accepted by physics community that
there is an agreement between the classical electromagnetism and the special
relativity (SR). Such an opinion is prevailing in physics already from
Einstein's first paper [1] on SR. The usual transformations of the
three-dimensional (3D) vectors of the electric and magnetic fields, $\mathbf{%
E}$ and $\mathbf{B}$ respectively (hereafter called the ``apparent''
transformations (AT)), are always considered to be the relativistically
correct Lorentz transformations (LT) of $\mathbf{E}$ and $\mathbf{B}$. (The
vectors in the 3D space will be designated in bold-face.) However, it is
recently proved [2] in the Clifford, i.e., geometric, algebra formalism that
the AT of $\mathbf{E}$ and $\mathbf{B}$ differ from the LT (boosts) of the
corresponding 4D quantities that represent the electric and magnetic fields.
The AT of $\mathbf{E}$ and $\mathbf{B}$ are first derived by Lorentz [3] and
Poincar\'{e} [4,5] and independently by Einstein [1] and subsequently
derived and quoted in almost every textbook and paper on relativistic
electrodynamics, e.g., [6] Eqs. (11.148) and (11.149). The proof (in the
tensor formalism) that the AT of $\mathbf{E}$ and $\mathbf{B}$ are not the
LT is given in [7] and in the geometric algebra formalism in [2] and [8] (a
more pedagogical version, on-line at: http:/fizika.phy.hr). The fundamental
difference is that in the AT, e.g., the components of the transformed $%
\mathbf{E}^{\prime }$ are expressed by the mixture of components of $\mathbf{%
E}$ and $\mathbf{B}$, and similarly for $\mathbf{B}^{\prime }$. However, the
correct LT always transform the 4D algebraic object representing the
electric field only to the electric field, and similarly for the magnetic
field, as in (\ref{bc}). The results from [7] and [2] are used to
investigate the LT and the AT of the Maxwell equations with $\mathbf{E}$ and
$\mathbf{B}$ in [9]. There it is shown that the Lorentz transformed Maxwell
equations are not of the same form as the original ones. This proves that,
contrary to the general opinion, the usual Maxwell equations are not
covariant under the LT but under the relativistically incorrect AT.

Comparisons with experiments, the motional emf [2], the Faraday disk [9] and
the Trouton-Noble experiment [10,11], show that the approach with 4D
geometric quantities always agrees with the principle of relativity and it
is in a true agreement (independent of the chosen inertial reference frame
and of the chosen system of coordinates in it) with experiments. This is not
the case with the usual approach in which the electric and magnetic fields
are represented by $\mathbf{E}$ and $\mathbf{B}$ that transform according to
the AT. (The name AT is introduced by Rohrlich [12] for the Lorentz
contraction; the Lorentz contracted length and the rest length are not
connected by the LT; they do not refer to the same 4D quantity.) In [13]
some well-known experiments, e.g., the Michelson-Morley type experiments,
are analyzed using Einstein's formulation of SR [1], which deals with the AT
(the AT of synchronously defined spatial length , i.e., the Lorentz
contraction and the AT of the temporal distance, i.e., the conventional
dilatation of time), and the new one which exclusively deals with 4D
geometric quantities. It is shown that all experiments that test SR are in a
true agreement with the geometric formulation. On the other hand the
agreement between the experiments that test SR and Einstein's formulation of
SR [1] is not a true agreement since it depends on the chosen
synchronization, e.g., Einstein's synchronization [1] or a drastically
different, nonstandard, ``radio'' synchronization. For different
synchronizations see, e.g., [14] and the second paper in [13], where both
synchronizations are used. \emph{This true agreement with experiments
directly proves the physical reality of the 4D geometric quantities.}

In this paper we shall present another, simple, but correct proof, that the
AT of $\mathbf{E}$ and $\mathbf{B}$ completely differ from the LT of the 4D
quantities that represent the electric and magnetic fields. Firstly Rosser's
derivation [15] of the AT of $\mathbf{E}$ and $\mathbf{B}$ will be presented
and then the objections to such derivation will be exposed. It will be shown
that the analogous derivation but with relativistically correct 4D
quantities lead to the correct LT of 4D geometric quantities $E^{a}$\textbf{%
\ }and\textbf{\ }$B^{a}$. Furthermore, the experimental searches for an
electric dipole moment (EDM) of a fundamental particle and their use of the
AT of $\mathbf{E}$ and $\mathbf{B}$ will be discussed.

\textit{Rosser's derivation} [15] \textit{of the AT of} $\mathbf{E}$ \textit{%
and} $\mathbf{B}$\textit{.} - A nice example of the derivation of the AT
that uses the 3D quantities is given in [15] Sec. 6.3. In that derivation it
is supposed that the usual expression for the Lorentz force as a 3D vector
must be of the same form in two relatively moving 4D inertial frames $S$ and
$S^{\prime }$, Eqs. (6.42) and (6.43) in [15]; in $S$,$\ \mathbf{F}_{L}%
\mathbf{=}q\mathbf{E}+q\mathbf{u}\times \mathbf{B}$ and in $S^{\prime }$,$\
\mathbf{F}_{L}^{\prime }\mathbf{=}q\mathbf{E}^{\prime }+q\mathbf{u}^{\prime
}\times \mathbf{B}^{\prime }$, where $\mathbf{u}$ and $\mathbf{u}^{\prime }$
are the 3D velocities of a particle. Then, in [15] Eqs. (1.53)-(1.55), the
usual transformations of components of $\mathbf{F}$ are presented, which we
write as $F_{x}^{\prime }=[F_{x}-(\beta _{u}/c)(\mathbf{Fu})]/(1-(\beta
_{u}u_{x}/c))$, $F_{y,z}^{\prime }=F_{y,z}/\gamma (1-(\beta _{u}u_{x}/c))$.
Substituting here the components of $\mathbf{F}_{L}$ and $\mathbf{F}%
_{L}^{\prime }$ and the components of $\mathbf{u}$ and $\mathbf{u}^{\prime }$%
, which are given by the usual transformations of the 3D velocity, e.g.,
Eqs. (1.26)-(1.28) in [15], or Eq. (11.31) in [6], Rosser derives the AT of
the components $E_{x,y,z}$ and $B_{x,y,z}$ of $\mathbf{E}$ and $\mathbf{B}$.
They are $E_{x}=E_{x}^{\prime }$, $E_{y}=\gamma (E_{y}^{\prime }+\beta
cB_{z}^{\prime })$, $E_{z}=\gamma (E_{z}^{\prime }-\beta cB_{y}^{\prime })$
and $B_{x}=B_{x}^{\prime }$, $B_{y}=\gamma (B_{y}^{\prime }-\beta
E_{z}^{\prime }/c)$, $B_{z}=\gamma (B_{z}^{\prime }+\beta E_{y}^{\prime }/c)$%
. These relations are Eqs. (6.40) and (6.41) in [15] or Eq. (11.148) in [6],
or the transformations derived in Sec.6 in [1]. Then $\mathbf{E}^{\prime }$
and $\mathbf{B}^{\prime }$ as \emph{geometric quantities in the 3D space}
are constructed in $S^{\prime }$ in the same way as in $S$, i.e.,
multiplying the spatial components $E_{x,y,z}^{\prime }$ and $%
B_{x,y,z}^{\prime }$ by \emph{the unit 3-vectors} $\mathbf{i}^{\prime },$ $%
\mathbf{j}^{\prime },$ $\mathbf{k}^{\prime }$. This yields the usual
transformations of $\mathbf{E}$ and $\mathbf{B}$\textbf{, }e.g., [15] Eqs.
(6.40a) and (6.41a), or [6] Eq. (11.149). Both the transformations for the
spatial components $E_{x,y,z}$, $B_{x,y,z}$ and for $\mathbf{E}$, $\mathbf{B}
$ are typical examples of the AT.

There are different shortcomings in the considered Rosser's derivation. They
are:

(i) The form invariance of the 3D Lorentz force doesn't follow from any
physical law; the principle of relativity doesn't say anything about the
form invariance of the 3D quantities. In [2] it is also shown that the form
invariance of the 3D Lorentz force doesn't agree with experiments on the
motional emf.

(ii) The forces $\mathbf{F}$, $\mathbf{F}^{\prime }$ are constructed from
the components $F_{x,y,z}$, $F_{x,y,z}^{\prime }$ and the unit 3-vectors $%
\mathbf{i}$, $\mathbf{j}$, $\mathbf{k}$, and $\mathbf{i}^{\prime }$, $%
\mathbf{j}^{\prime }$, $\mathbf{k}^{\prime }$ respectively. There are not
either the LT or the AT which transform the unit 3-vectors $\mathbf{i}$, $%
\mathbf{j}$, $\mathbf{k}$ into the unit 3-vectors $\mathbf{i}^{\prime }$, $%
\mathbf{j}^{\prime }$, $\mathbf{k}^{\prime }$. Consequently $\mathbf{F}%
_{L}^{\prime }$ \emph{is not obtained by the LT from} $\mathbf{F}_{L}$.

(iii) In the geometric approach the physical meaning is attributed only to
the 4D geometric quantities. The LT always correctly transform the whole 4D
quantity and they do not refer to some parts of 4D quantities like
components of $\mathbf{F}$. This means that the transformations of
components of $\mathbf{F}$ are not well-defined in the 4D spacetime and they
are not the LT but the AT.

The same objections hold for the well-known transformations of $\mathbf{u}$.

Furthermore, the above AT of $F_{x,y,z}$ are derived supposing that the
``relativistic'' equations of motion have the same form in two relatively
moving inertial frames $S$ and $S^{\prime }$, i.e., $\mathbf{F}=d\mathbf{p}%
/dt$ in $S$ and $\mathbf{F}^{\prime }=d\mathbf{p}^{\prime }/dt^{\prime }$ in
$S^{\prime }$, where $\mathbf{p=}m\gamma _{u}\mathbf{u}$, $\gamma
_{u}=(1-\left| \mathbf{u}\right| ^{2}/c^{2})^{-1/2}$, see, e.g., Eqs. (1.39)
and (1.40) in [15], or Sec. 12.2 and 12.3 (with the Lorentz force) in [6].

However, as shown in, e.g., [16], \emph{a 3D quantity cannot correctly
transform under the LT, which means that it does not have an independent
physical reality in the 4D spacetime; it is not the same quantity for
relatively moving observers in the 4D spacetime}. Hence, it is not true that
the the above equations with 3-vectors are the relativistic equations of
motion since the primed 3D quantities are not obtained by the LT from the
unprimed ones, but they are obtained in terms of the AT of $\mathbf{F}$ and $%
\mathbf{p}$.

Instead of the equations with $\mathbf{F}$ and $\mathbf{p}$ one has to use
the equation of motion with 4D geometric quantities, $K^{a}=dp^{a}/d\tau $,$%
\ p^{a}=mu^{a}$, where $p^{a}$ is the proper momentum (4-vector), $\tau $ is
the proper time and $K^{a}$ is a 4D force (4-vector), e.g., the Lorentz
force $K_{L}^{a}$ as a 4-vector, which is given below. The 4-vectors $K^{a}$%
, $p^{a}$, $u^{a}$ are correctly defined quantities in the 4D spacetime,
both theoretically and \emph{experimentally}. They are defined without
reference frames. Latin indices a,b,c,d are to be read according to the
abstract index notation, as in [17] and [13,14,18]. When some basis has been
introduced (here the standard basis $\left\{ \gamma _{\mu }\right\} $) then,
e.g., $K^{a}$, is represented in that basis as $K^{a}=K^{\mu }\gamma _{\mu }$%
, where $\gamma _{\mu }$ are the basis 4-vectors and $K^{\mu }$ are the
components; Greek indices run from 0 to 3 and they denote the components of
the geometric object, $K^{a}$. The $\{\gamma _{\mu }\}$ basis corresponds to
the Einstein system of coordinates; the Einstein synchronization [1] of
clocks and Cartesian spatial coordinates.

In contrast to awkward transformations of the components of $\mathbf{F}$,
and the similar ones for $\mathbf{u}$, the LT of the components $K^{\nu }$
of the 4-vector $K^{a}$ are very simple
\begin{equation}
K^{\prime 0}=\gamma (K^{0}-\beta K^{1}),\ K^{\prime 1}=\gamma (K^{1}-\beta
K^{0}),\ K^{\prime 2,3}=K^{2,3}  \label{ka}
\end{equation}
(for the boost in the $\gamma _{1}$ direction). The same holds for the
components of $p^{a}$ and $u^{a}$. Observe that $K^{a}$ is the same quantity
for relatively moving 4D observers, $K^{a}=K^{\mu }\gamma _{\mu }=K^{\prime
\mu }\gamma _{\mu }^{\prime }$. The components $K^{\mu }$ transform by the
LT (\ref{ka}) and the basis 4-vectors $\gamma _{\mu }$ transform by the
inverse LT thus leaving the whole 4D quantity\ invariant under the passive
LT. It is completely different than in the AT of the components of $\mathbf{F%
}$.

\textit{The derivation of the LT of }$E^{a}$ \textit{and} $B^{a}$\textit{.}
- Let us apply the same procedure as in Rosser's derivation [15] of the AT
of the components of $\mathbf{E}$ and $\mathbf{B}$, but now with
well-defined 4D quantities, the 4D Lorentz force, $%
K_{L}^{a}=(q/c)F^{ab}u_{b} $, and $u^{b}$, the 4-velocity of a charge $q$.
It will be shown that the result of such procedure will be the LT of the
4-vectors $E^{a}$, $B^{a}$, i.e., of their components $E^{\mu }$, $B^{\mu }$
and not the AT of $E_{x,y,z} $, $B_{x,y,z}$. The decomposition of $F^{ab}$
in terms of $E^{a}$ and $B^{a}$, see, e.g., [14], [9], is given by

\begin{eqnarray}
F^{ab} &=&(1/c)(E^{a}v^{b}-E^{b}v^{a})+\varepsilon ^{abcd}v_{c}B_{d},  \notag
\\
E^{a} &=&(1/c)F^{ab}v_{b},\ B^{a}=(1/2c^{2})\varepsilon ^{abcd}F_{bc}v_{d}.
\label{f}
\end{eqnarray}
The $E^{a}$ and $B^{a}$ are the electric and magnetic field 4-vectors
measured by an observer moving with 4-velocity $v^{a}$ in an arbitrary
reference frame, $v^{a}v_{a}=c^{2},$ and it holds that $%
v_{a}E^{a}=v_{b}B^{b}=0$. Note that $E^{a}$ and $B^{a}$ depend not only on $%
F^{ab}$ but on $v^{a}$ as well. The frame of ``fiducial''\ observers is the
frame in which the observers who measure $E^{a}$ and $B^{a}$ are at rest.
That frame with the standard basis $\{\gamma _{\mu }\}$ in it is called the $%
\gamma _{0}$-frame. In the $\gamma _{0}$-frame $v^{a}=c\gamma _{0}$, which,
with (\ref{f}), yields that $E^{0}=B^{0}=0$ and $E^{i}=F^{i0}$, $%
B^{i}=(1/2c)\varepsilon ^{ijk0}F_{jk}$. Thus, in the $\gamma _{0}$-frame
only spatial components $E^{i}$ and $B^{i}$ remain and they correspond to
the components of $\mathbf{E}$ and $\mathbf{B}$. The equation (\ref{f})
indicates that $F^{ab}$ can be taken as the primary quantity for the whole
electromagnetism. $E^{a}$ and $B^{a}$ are then derived from $F^{ab}$ and $%
v^{a}$. Such formulation, but in the geometric algebra formalism, is
presented in [10]. The Lorentz invariant field equations with $E^{a}$ and $%
B^{a}$ (i.e., with 1-vectors $E$ and $B$) are presented in [14], [9] (i.e.,
in [9], [19]). Such field equations but with the complex combination of $%
E^{a}$ and $B^{a}$ (i.e., of $E$ and $B$), the 4D Majorana form, and also
the Dirac-like equation for the free photon are given in [14] (i.e., in
[19]).

Substituting the decomposition of $F^{ab}$ (\ref{f}) into the expression for
the 4D Lorentz force $K_{L}^{a}=(q/c)F^{ab}u_{b}$ one finds $%
K_{L}^{a}=(q/c^{2})[(v^{b}u_{b})E^{a}+\varepsilon
^{bacd}v_{b}u_{c}cB_{d}-(E^{b}u_{b})v^{a}]$. When $K_{L}^{a}$ is written in
the $\left\{ \gamma _{\mu }\right\} $ basis then it becomes
\begin{equation}
K_{L}^{a}=(q/c^{2})[(v^{\nu }u_{\nu })E^{\mu }+\varepsilon ^{\lambda \mu \nu
\rho }v_{\lambda }u_{\nu }cB_{\rho }-(E^{\nu }u_{\nu })v^{\mu }]\gamma _{\mu
}.  \label{kf}
\end{equation}
Let us take that the $S$ frame is the $\gamma _{0}$ - frame in which $v^{\mu
}=(c,0,0,0)$. Then, using (\ref{ka}), we start with the component $%
K_{L}^{\prime 2}$ for which it holds that $K_{L}^{\prime 2}=K_{L}^{2}$. This
equation corresponds to Eq. (6.44) in Rosser's derivation [15]. Substituting
$K_{L}^{2}$ and $K_{L}^{\prime 2}$, which are determined by (\ref{kf}), into
$K_{L}^{\prime 2}=K_{L}^{2}$, that equation becomes
\begin{equation}
cu_{0}E^{2}-c^{2}u_{1}B_{3}+c^{2}u_{3}B_{1}=cu_{0}E^{\prime
2}-c^{2}u_{1}B_{3}^{\prime }+c^{2}u_{3}\gamma (B_{1}^{\prime }+\beta
B_{0}^{\prime }),  \label{c}
\end{equation}
where it is used that $v^{\prime \mu }=(\gamma c,-\gamma \beta c,0,0)$ (the
``fiducial''\ observers are moving in $S^{\prime }$) and the components $%
u^{\prime \mu }$ are determined by the LT from $u^{\mu }$, which are the
same as (\ref{ka}). The equation (\ref{c}) corresponds to Eq. (6.46) in
Rosser's derivation [15]. Then, as in [15], Eq. (\ref{c}) must be valid
whatever the value of $u^{\mu }$. If $u^{\mu }$ can have various values,
then the terms containing $u_{0}$, $u_{1}$ and $u_{3}$ on the left-hand side
of (\ref{c}) must be equal to the terms containing $u_{0}$, $u_{1}$ and $%
u_{3}$ on the right-hand side of (\ref{c}) respectively. This leads to the
equations
\begin{equation}
E^{2}=E^{\prime 2},\quad B^{1}=\gamma (B^{\prime 1}+\beta B^{\prime 0}),\
B^{3}=B^{\prime 3}.  \label{bc}
\end{equation}
Proceeding in the same way one can get the LT for all components $E^{\mu }$
and $B^{\mu }$, which are the same as the LT (\ref{ka}). Observe that the
transformations of components in (\ref{bc}) are completely different than
the AT of the corresponding components $E_{y}$, $B_{x}$ and $B_{z}$. The
same fundamental difference would be found for all other components. This is
very simple and illustrative proof of the existence of the fundamental
difference between the LT of the 4D quantities that represent the electric
and magnetic fields and the AT of $\mathbf{E}$ and $\mathbf{B}$.

The obtained results will significantly influence the interpretations of
measurements of an EDM, e.g., [20], [21]. In all experimental searches for a
permanent EDM of particles the AT of $\mathbf{E}$ and $\mathbf{B}$ are
frequently used and considered to be relativistically correct; i.e., that
they are the LT of $\mathbf{E}$ and $\mathbf{B}$. Thus, in a recent new
method of measuring EDMs in storage rings [20] the so-called motional
electric field is considered to arise ``according to a Lorentz
transformation''\ from a vertical magnetic field $\mathbf{B}$ that exists in
the laboratory frame; $\mathbf{E}^{\prime }=\gamma c\mathbf{\beta \times B}$%
. That field $\mathbf{E}^{\prime }$ plays a decisive role in the mentioned
new method of measuring EDMs. Note that the field $\mathbf{E}^{\prime }$ is
in the rest frame of the particle $S^{\prime }$ but the measurement of EDM
is in the laboratory frame $S$. Similarly happens in [21] and many others in
which 'motional magnetic field' $\mathbf{B}^{\prime }=(\gamma /c)\mathbf{%
E\times \beta }$ appears in the particle's rest frame as a result of the AT
of the $\mathbf{E}$ field from the laboratory. The same interpretation with
the AT of $\mathbf{E}$ and $\mathbf{B}$ appears when the quantum phase of a
moving dipole is considered, e.g., [22]. For example, when the R\"{o}ntgen
phase shift is considered then it is asserted in the second paper in [22]
that in ``the particle rest frame the magnetic flux density $\mathbf{B}$ due
to the magnetic line is perceived as an electric field''\ $\mathbf{E}%
^{\prime }=\mathbf{v}\times \mathbf{B}$. This is objected in [23].

However, the transformations of $\mathbf{E}$ and $\mathbf{B}$ are not the LT
but the AT. They have to be replaced by the LT of $E^{a}$ and $B^{a}$. Then,
the LT of components of a 4-vector, as in (\ref{ka}), transform $B^{\mu }$
from $S$ again to $B^{\prime \mu }$ in $S^{\prime }$ and similarly for $%
E^{\mu }$; \emph{there is no mixing of components.} Thus, in this 4D
geometric approach, there is no induced $\mathbf{E}^{\prime }$ as in [20]
and [22], and there is no 'motional magnetic field' $\mathbf{B}^{\prime }$
as in [21]. As shown in [18], and particularly in [24], the same happens
with dipole moments $\mathbf{d}$ and $\mathbf{m}$ and the corresponding
4-vectors $d^{a}$ and $m^{a}$.

\textit{Conclusions.} - This proof of the fundamental difference between the
AT and the LT of the electric and magnetic fields strongly indicates that
the experimentalists who search for an EDM, e.g., [20] and [21], and, e.g.,
those who observe the Aharonov-Casher phase shift, [25], will need to
reexamine the results of their measurements taking into account that the
transformations of $\mathbf{E}$, $\mathbf{B}$, $\mathbf{d}$, $\mathbf{m}$
are not the LT but the AT and replacing these AT by the LT of the 4D
geometric quantities $E^{a}$, $B^{a}$, $d^{a}$, $m^{a}$. In our 4D spacetime
an independent physical reality (in Minkowski's sense) has to be attributed
to the 4D geometric quantities and the whole physics would need to be
expressed by such quantities and not, as generally accepted, by the
relativistically incorrect 3D quantities with their AT.\medskip \bigskip

\noindent \textbf{References\medskip }

\noindent \lbrack 1] A. Einstein, Ann. Physik. \textbf{17}, 891 (1905), tr.
by W. Perrett and G.B. Jeffery, in \textit{The Principle of Relativity}

(Dover, New York, 1952).

\noindent \lbrack 2] T. Ivezi\'{c}, Found. Phys. Lett. \textbf{18}, 301
(2005).

\noindent \lbrack 3] H.A. Lorentz, Proceedings of the Academy of Sciences of
Amsterdam \textbf{6,} (1904), tr. by W. Perrett and G.B.

Jeffery, in \textit{The Principle of Relativity }(Dover, New York, 1952).

\noindent \lbrack 4] H. Poincar\'{e}, Rend. del Circ. Mat. di Palermo
\textbf{21,} 129 (1906). See also two fundamental Poincar\'{e}'s papers with

notes by Logunov.

\noindent \lbrack 5] A.A. Logunov, Hadronic J. \textbf{19,} (1996) 109.

\noindent \lbrack 6] J.D. Jackson, \textit{Classical Electrodynamics}
(Wiley, New York, 1998) 3rd edn.

\noindent \lbrack 7] T. Ivezi\'{c}, Found. Phys. \textbf{33}, 1339 (2003)%
\textbf{. }

\noindent \lbrack 8] T. Ivezi\'{c}, Fizika A (Zagreb) \textbf{17}, 1 (2008).

\noindent \lbrack 9] T. Ivezi\'{c}, Found. Phys. \textbf{35}, 1585 \textbf{(}%
2005\textbf{)}.

\noindent \lbrack 10] T. Ivezi\'{c}, Found. Phys. Lett. \textbf{18,} 401
(2005).

\noindent \lbrack 11] T. Ivezi\'{c}, Found. Phys. \textbf{37,} 747 (2007).

\noindent \lbrack 12] F. Rohrlich, Nuovo Cimento\ B, \textbf{45}, 76 (1966).

\noindent \lbrack 13] T. Ivezi\'{c}, Found. Phys. Lett. \textbf{15,} 27
(2002); physics/0103026.

\noindent \lbrack 14] T. Ivezi\'{c}, Found. Phys. \textbf{31}, 1139 (2001).

\noindent \lbrack 15] W.G.W. Rosser, \textit{Classical Electromagnetism via
Relativity} (Plenum, New York, 1968).

\noindent \lbrack 16] T. Ivezi\'{c}, Found. Phys. \textbf{36}, 1511 (2006)%
\textbf{; }Fizika A (Zagreb) \textbf{16}, 207 (2007).

\noindent \lbrack 17] R.M. Wald, \textit{General Relativity} (Chicago
University, Chicago, 1984).

\noindent \lbrack 18] T. Ivezi\'{c}, Phys. Rev. Lett. \textbf{98}, 108901
(2007).

\noindent \lbrack 19] T. Ivezi\'{c}, EJTP \textbf{10}, 131 (2006).

\noindent \lbrack 20] F. J. M. Farley et al., Phys. Rev. Lett. \textbf{93},
052001 (2004).

\noindent \lbrack 21] A. L. Barabanov, R. Golub and S. K. Lamoreaux, Phys.
Rev. A \textbf{74}, 052115 (2006); C. A. Baker et al., Phys.

Rev. Lett. \textbf{97}, 131801 (2006); B. C. Regan, E. D. Commins, C. J.
Schmidt, D. DeMille, Phys. Rev. Lett. \textbf{88,}

071805 (2002).

\noindent \lbrack 22] M. Wilkens, Phys. Rev. Lett. \textbf{72}, 5 (1994); S.
A. R. Horsley, M. Babiker, Phys. Rev. Lett. \textbf{95}, 010405 (2005).

\noindent \lbrack 23] T. Ivezi\'{c}, Phys. Rev. Lett. \textbf{98}, 158901
(2007).

\noindent \lbrack 24] T. Ivezi\'{c}, arXiv: 0801.0851v2 [physics.gen-ph].

\noindent \lbrack 25] A. Cimmino, et al., Phys. Rev. Lett. \textbf{63}, 380
(1989); K. Sangster, et al., Phys. Rev. A \textbf{51}, 1776 (1995).

\end{document}